\documentstyle[12pt,psfig]{article}

\makeatletter
\def\ps@myheadings{\let\@mkboth\@gobbletwo
 \def\@oddhead{\hfil\rightmark}%
 \def\@oddfoot{\hfil\rm\thepage\hfil}%
 \def\@evenhead{\@oddhead}%
 \def\@evenfoot{\@oddfoot}\def\sectionmark##1{}\def\subsectionmark##1{}}
\makeatother

\author{\large  Matteo Bertolini$^1$, Pietro Fr\`e$^2$, Roberto Iengo$^1$, 
Claudio A. Scrucca $^{1,3}$\\ \\
{\normalsize \em $^1$ International School for Advanced Studies and INFN, 
V. Beirut 2-4, 34013 Trieste, Italy}\\
{\normalsize \em $^2$ Dipartimento di Fisica teorica di Torino and INFN, 
V. P. Giuria 1, 10100 Torino, Italy}\\
{\normalsize \em $^3$ Lud.-Max.-Univ., Sektion Physik, Univ. M\"unchen, 
Theresienstr. 37, M\"unchen 2, Germany}}

\title{
\vskip -30pt
\normalsize
\begin{flushright}
SISSA REF 112/98/EP
\end{flushright}
\vskip 20pt
\Huge D3-branes dynamics and black holes}

\date{}

\begin{document}

\maketitle

\thispagestyle{myheadings}

\begin{abstract}

Using the D3-brane as the fundamental tool, we adress two aspects of D-branes physics. 
The first regards the interaction between two electromagnetic dual D-branes in 10 
dimensions. In particular, we give a meaning to {\it both} even and odd spin structure 
contributions, the latter being non vanishing for non zero relative velocity $v$ (and 
encoding the Lorentz-like contribution). The second aspect regards the D-brane/black holes 
correspondence. We show how the 4 dimensional configuration corresponding to 
a {\it single} D3-brane wrapped on the orbifold $T^6/Z_3$ represents a regular Reissner-
Nordstr\"om  solution of $d=4\;N=2$ supergravity

\end{abstract}

\begin{center}

Talk presented by Matteo Bertolini

\end{center}
\section{Introduction and Summary}
In the last few years the study of D-branes \cite{pol} and their dynamics has revealed to be, 
under many respects,  one of the most promising aspects of string theory to be investigated. 
In this contribution I will adress two different aspects regarding D-brane physics. 

Using the D3-brane as the fundamental tool I will first of all consider the interaction of two 
moving D3-branes in 10 dimensions. The fact that the D3-brane is a dyon (namely charged both 
electrically and magnetically with respect to the R-R 4-form) implies that the gauge interaction 
of two of these objects has both a Coulomb-like {\it and} a Lorentz-like contribution. From string 
theory point of view the two are encoded respectively in the even and the odd Ramond-Ramond spin 
structures emerging after GSO projection on the relevant cylinder amplitude. The treatment of the 
odd spin structure is delicate: a naive computation would give back always a vanishing result 
because of the presence of at least two fermionic zero-modes coming from transverse directions.
This is not peculiar of the D3-brane of course, but it is a general problem affecting the 
interaction of any couple of electromagnetic dual $p$ and $(6-p)$ D-branes. The first result, 
therefore, will be to illustrate a way to treat the non-trivial odd spin structure 
contribution in order to get the correct phase-shift we would expect from a field theory 
point of view.

In the second part of my contribution I will consider the 4 dimensional configuration corresponding 
to a D3-brane wrapped on the orbifold $T^6/Z_3$. Once integrated over the compact coordinates, 
this configuration corresponds to an exact black hole solution of the effective $N=2$ 
supergravity theory in 4 dimensions. More precisely, and this is the most interesting fact, 
it represents an extremal dyonic Reissner-Nordstr\"om (R-N) black hole with non vanishing 
entropy. This a particular example of what happens in generic Calabi-Yau (CY) compactifications 
($T^6/Z_3$ is an orbifold limit of a CY): as opposite to compactifications on tori, in 
CY compactifications one can get regular solutions, in 4 dimensions, even with single 
charged objects. The non-trivial topological structure of CY supersymmetric cycles on 
which the D-branes are wrapped can ``re-gularize'' the solution, as the intersection of 
different D-branes does on tori. The D-brane/black hole correspondence has been investigated 
both from a string and supergravity point of view and results are shown to agree. Moreover, it 
can be shown how the actual values of the 4 dimensional electric and magnetic charges depend 
explicitly on branes' orientation in the compact space. 

\section{The interaction of two D3-branes in 10 dimensions}

As well known, D-branes can be defined as hypersurfaces on which open strings can end and 
therefore the potential between two interacting D-branes is expressed by vacuum fluctuations 
of open strings streched between them. However, thanks to the conformal invariance of the string 
world sheet, one can study D-branes interactions also from closed string point of view. In this 
latter case the D-brane can be thought as a source of closed strings and the interaction is 
mediated by exchange of closed string states.
A very appropriate tool in order to describe D-branes and their dynamics from a closed strings 
point of view is the {\it boundary state formalism} \cite{polcai}. A boundary state $|B \rangle$ 
describing a D-brane can be defined as a coherent state written in terms of closed string 
oscillators which implement the boundary conditions (N or D) of strings which the brane can 
emit. The typical structure of a boundary state describing a D-brane is of the following form: 
$$
|B \rangle\;=\;|B \rangle_{bos} |B \rangle_{ghost} |B \rangle_{fer} |B \rangle_{superghost}
\;\simeq\;|B \rangle_{0}\; \mbox{exp}\left[ \sum_{n\geq 1} 
\frac 1n \left(\alpha^\mu_{-n} S_{\mu\nu}\tilde{\alpha}^\nu_{-n} + ...\right)\right] |0\rangle
$$
where there are contributions both from physical and ghost oscillators and where $|B \rangle_{0}\,$ 
encodes the zero modes part of the boundary state. The matrix $S_{\mu\nu}$ is a diagonal matrix 
with $+1$ or $-1$ entries according to Dirichelet or Neumann nature of the corresponding 
directions, respectively.
 
Let us now consider the 10 dimensional interaction of two D3-branes moving with velocities 
$V_1=tanhv_1$ and $V_2=tanhv_2$ along $X^1$ direction, tilted by angles $\theta^a_1\,,\,\theta^a_2$ 
on the 3 planes $X^a,X^{a+1}$ ($a=4,6,8$), that will eventually become compact, and 
with transverse positions $\vec Y_1\,,\,\vec Y_2$:

\vskip 20pt
\input epsf
\epsfysize=135pt
\centerline{\epsffile{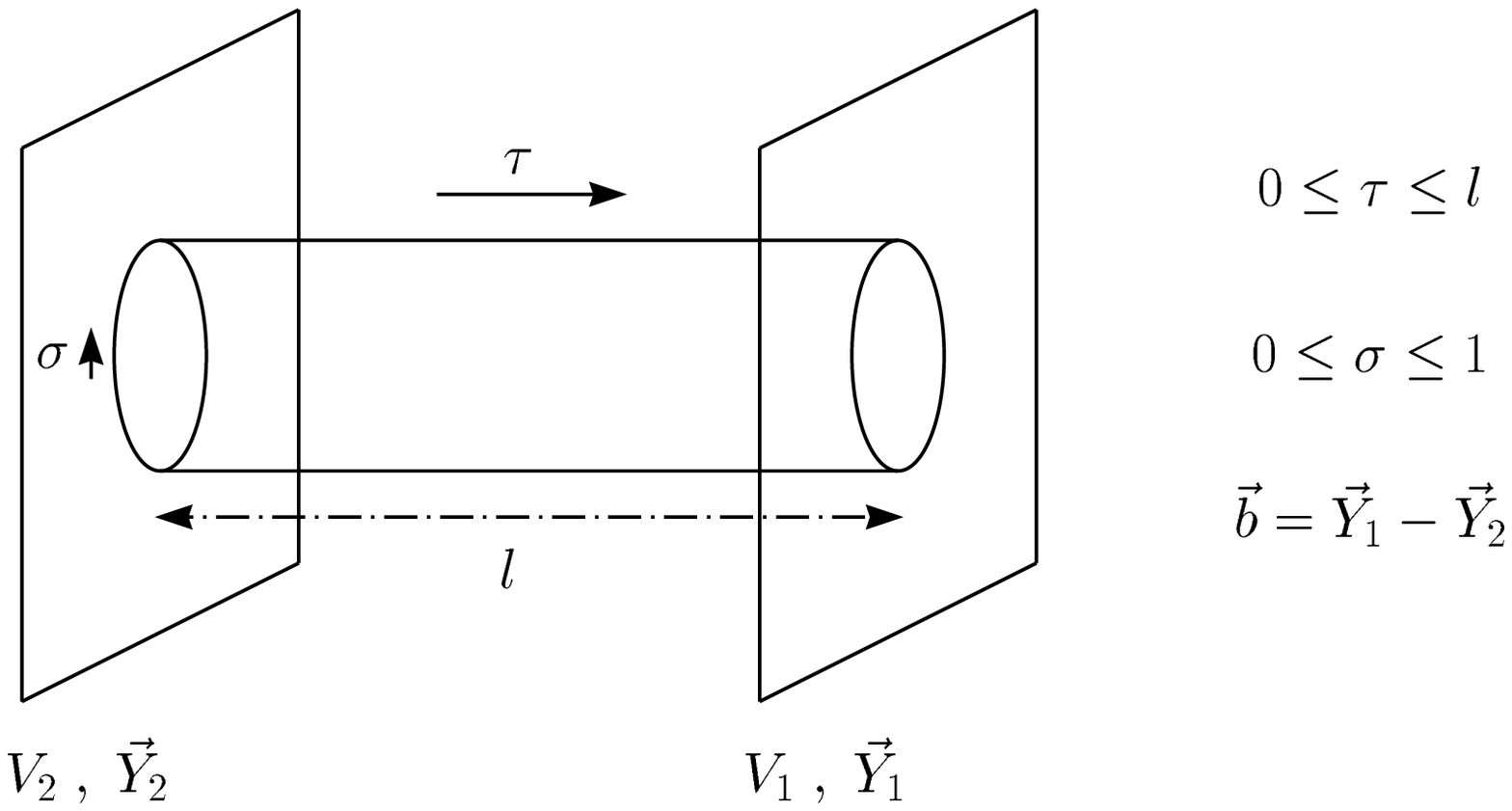}}

The parameter $l$ is the world sheet distance while $\vec b$ is the physical impact 
parameter. 

In the boundary state formalism the interaction is given by the correlation between 
two boundary states and in our case can then be written in the following way:
\begin{equation}
\label{amp}
{\cal A} =\frac{\mu^2}{16} \sum_s
<v_1,\vec Y_1,\theta^1_a|\int_0^{\infty} dl e^{-lH}
|v_2,\vec Y_2, \theta^2_a>_s
\end{equation}
$\mu=\sqrt{2\pi}$ is the electric (and magnetic) density charge of the D3-brane while the sum 
over {\it s} is made on the 4 spin structures emerging after GSO projection on the cylinder. 
The boun-dary state $|B,v\rangle$ describing a moving brane is obtained from the static one 
through a Lorentz transformation with velocity 
$v$: $| B, v \rangle = exp ( -ivJ^{01})| B \rangle$ \cite{dive}.

The essential kinematic is encoded in the bosonic zero-modes contribution which can be written 
as a product of delta functions enforcing the boundary conditions for the center of mass 
position operator $X_o^\mu$, that is a Fourier superposition of momentum states. Hence, in 
momentum space, the boundary states for the two branes are:
\begin{eqnarray}
|B,v_1,\theta^1_a,\vec Y_1\rangle \,= \, \int \frac{d^{6}\vec k}{(2\pi)^{6}}
e^{i \vec k_B \cdot \vec Y_1} |v_1,\theta^1_a\rangle \otimes|k_B\rangle \;,\; 
|B,v_2,\theta^2_a,\vec Y_2\rangle \,= \, \int \frac{d^{6}\vec q}{(2\pi)^{6}}
e^{i \vec q_B \cdot \vec Y_2} |v_2,\theta^2_a\rangle \otimes|q_B\rangle \nonumber
\end{eqnarray}
with $\vec k_B$ and  $\vec q_B$ being the boosted and rotated momenta:
\begin{eqnarray}
&& k_B^\mu = (\sinh v_1 k^1, \cosh v_1 k^1, k^2, k^3, \cos \theta^1_a k^a, \sin \theta^1_a k^a)
\nonumber \\
&& q_B^\mu = (\sinh v_2 q^1, \cosh v_2 q^1, q^2, q^3, \cos \theta^2_a q^a, \sin \theta^2_a q^a)
\nonumber
\end{eqnarray}
and $k^2,k^3,q^2$ and $q^3$ the momenta along the only two completely transverse directions.

Integrating over momenta and taking into account momentum conservation which for 
non-vanishing $v \equiv v^{(1)} - v^{(2)}$ and $\theta_a \equiv \theta_a^{(1)} - \theta_a^{(2)}$
forces all the Dirichlet momenta but $k^2, k^3$ to be zero, the amplitude (\ref{amp}) reads:
\begin{equation}
\label{amp10}
{\cal A}=\frac {\mu^2}{2\sinh |v| \prod_a 2\sin |\theta_a|} \int_{0}^{\infty}
\frac {dl}{4\pi l} e^{-\frac {b^2}{4 l}} \sum_\alpha Z_B Z^\alpha_F
\end{equation}
where $Z^s_{B,F}$ are the bosonic and fermionic partition functions:
$$
Z^s_{B,F}=<v^{(1)},\theta_a^{(1)}|e^{-lH}|v^{(2)},\theta_a^{(2)}>^s_{B,F}
$$ 
In the above expression, only the oscillator modes of the string coordinates $X^\mu$
appear, since we have already integrated over the center of mass coordinate. 

In order to compare string and supergravity results, we take in (\ref{amp10}) the limit of 
large impact parameter ($b\rightarrow\infty$) . In this limit only world sheets with 
$l\rightarrow \infty$ contribute and the behaviour of partition functions simplifies a lot: 
\begin{eqnarray}
\label{spin}
&& Z_B \rightarrow 1 \;, \nonumber \\
&& Z_F^{even} \rightarrow 2 \cosh v \prod_a 2 \cos \theta_a  
- 2 (2 \cosh 2v + \sum_a 2\cos 2 \theta_a ) \nonumber \\
&& Z_F^{odd} \rightarrow 2 \sinh v \prod_a 2 \sin \theta_a \cdot 0 
\end{eqnarray}
If the two branes are parallel and at rest, i.e. if $\theta_a=0$ and $v=0$, the configuration 
is BPS and one gets $Z_F^{even}=0$ that expresses the well known no-force condition between 
two BPS states. If the two branes are parallel but not at rest, i.e. if $\theta_a=0$ but 
$v\not= 0$, one gets $Z_F^{even}\sim v^4$ that is infact the phase shift leading order 
contribution for moving D-branes. Finally, in the more general case, the configuration is not 
BPS and one has $Z_F^{even}\not= 0\;$. However, for suitable angles ($\sum_a \theta_a=2\pi n$) 
the configuration can still be BPS, although preserving less supersymmetry. This can be seen 
looking to the low velocity dependence of the partition functions; indeed in this case 
$Z_F^{even}\sim v^2$.   

It seems to be no contribution from $ Z_F^{odd}$ because of the $0$ always present in its 
expression. That $0$ comes from the fermionic zero-modes of the only two completely transverse 
directions left, $X^2$ and $X^3$, and has to be soaked up if one wants to give a meaning to 
the odd spin structure. In the Ramond sector, odd and even spin structures are responsible 
for the gauge interaction between the two branes (the NS sector encodes the gravitational 
contribution). D3-branes are dyons and therefore, for relative velocity $v \not= 0$ 
there is both, in general, an electric (Coulomb-like) and a magnetic (Lorentz-like) interaction. 
The former is encoded in the even spin structure while the latter in the odd one. This means that 
one would expect a non vanishing contribuiton also from the odd spin structure partition 
function (and the same must hold for any couple of electromagnetic dual D-branes).
One can naively understand this point looking to the tensorial structure of the two (even and odd) 
RR contributions. From (\ref{spin}) one sees that:

\begin{equation}
\label{even}
<v^1,\theta^a_1|e^{-lH}|v^2,\theta^a_2>_{RR+} \sim \cosh v \, \prod_a \,\cos \theta_a 
\nonumber 
\end{equation}
while
\begin{equation}
\label{odd}
<v^1,\theta^a_1|[...] e^{-lH}|v^2,\theta^a_2>_{RR-} \sim \sinh v \,\prod_a \, 
\sin \theta_a \nonumber 
\end{equation}
The cosine is a signal of a radial (Coulomb) force while the sine of an orthogonal (Lorentz) one.  
The simbol $[...]$ in the odd spin structure expression indicates the insertion of suitable 
supercurrents. In \cite{bis} it has been shown infact that with this insertion one can reproduce 
in string theory the expected field theory result. Namely, togheter with the primary necessity 
of soaking up fermionic zero-modes, one gets back, as a by-product, the right tensorial structure 
of the interaction with respect to the transverse non-compact directions, namely the exterior 
product structure that is charateristic of a Lorentz force. Notice that in order to get a non 
vanishing contribution from the odd spin structure one also needs the two 3-branes being non 
parallel (i.e. $\theta_a\not= 0$). This is again quite obvious: the condition for having non 
zero Lorentz interaction between {\it extended} objects is to have a complete non parallelism 
between the corresponding world-volume. This is a generalization of what happens in 4 dimensions 
to point-like objects where the Lorentz interaction is non vanishing for non zero relative 
velocity, this conditon being rephrased saying that there is a tilting between the worldlines 
of the two particles.  

An essentially analogous result has been achieved in a T-dual situation, namely a D0-brane 
moving in the background of a D6-brane \cite{d0d6}. With the insertion of suitable regulator 
for matter and superghost zero modes it has been found  a contribuiton from the odd spin 
structure as a scale-independent, velocity-dependent potential $V_{06}(r) = - \frac {v}{2r}$ 
whose {\it a posteriori} interpretation is of a Lorentz-like potential. In this way the direct 
result is not that of a phase-shift, however the final essential conclusion does not change, 
namely that the odd spin structure encodes the magnetic interaction contribution in a scattering 
amplitude of a pair of electromagnetic dual $p$ and $(6-p)$ D-branes and is in general 
different from zero. 

\section{The black hole configuration in 4 dimensions}

Let us now consider the four dimensional configuration corresponding to a D3-brane wrapped on a 
particular 6-dimensional compact space, the $T^6/Z_3$ orbifold. 
What has been acheived in the last few years about the D-brane/black hole correspondence is 
essentially that various D-branes configurations can in general give a microscopic description 
of both extremal and non-extremal black holes emerging in 4 dimensions by compactification of 
p-branes solutions of type II supergravity in 10 dimensions.

Starting from type II theory in 10 dimensons, for toroidal compactifications one obtains an 
effective $N=8$ supergravity in 4 dimensions while for compactifications on more generic 
CY spaces one ends up with $N=2$ theory. Configurations for which a microscopic 
description can be given correspond mainly to toroidal compactifications of ``intersecting 
D-branes''\cite{bala}. In $N=2$ compactifications, from microscopic point of view, much less 
has been said, the essential reason being that for curved D-branes Polchinski's prescription 
is more difficult to be implemented. The orbifold $T^6/Z_3$ is actually a limit of a CY 
space (with Hodge numbers $h_{(1,1)}=9$ and $h_{(1,2)}=0$) and so falls in this latter class. 
It's utility resides in the fact that it has a sufficently simple structure to be treated with 
usual boundary state techniques but gives, on the other hand, sufficently interesting results. 
Indeed, what I will show is that in our case, as in many other CY compactifications, 
one can obtain regular black hole solutions even with less than 4 charged objects, as opposite 
to the toroidal case. The intuitive reason for that is that the non-trivial topological structure 
of CY space implies that a single D-brane can ``intersect'' with itself on a given supersymmetric 
cycle, therefore mimicking the actual intersection of different D-branes needed on tori in order 
to get regular solutions.

Strictly speaking a R-N black hole is a non singular spherical solution 
of Maxwell-Einstein gravity that however can be consistently extended to be $N=2$ supersymmetric. 
More in general, in matter coupled $N=2$ supergravity, one can have generalized regular solutions 
which turn out to be R-N near the horizon, that is with an $AdS_2 \times S^2$ topology. In 
particular, one have the so called {\it double extreme} solution when the vector multiplets 
scalars are taken to be constant and equal to their fixed values they anyhow must get at the 
horizon \cite{fer}. The hypermultiplet scalars, on the other hand, couple minimally to the gauge 
fields and can therefore always taken to be neutral in a given solution.

When one compactifies type IIA or IIB  supergravity on a CY manifold, the number of vector 
and hypermultiplets of the relevant $N=2$ theory is dictated by the two relevant Hodge numbers 
($h_{(1,1)}$,$h_{(1,2)}$) charaterizing the C-Y space. In our case, that is type IIB on $T^6/Z_3$, 
one has $0$ vector multiplets and $10$ hypermultiplets. Therefore there are not vector multiplet 
scalars and the solution is automatically double extreme. From a supergravity point of view the 
solution is therefore quite straightforward. Let us now analize the configuration both from a 
macroscopic and a microscopic point of view.

A type II $p$-brane usually couples to the metric, the dilaton and the corresponding (p+1) gauge 
potential. The peculiar property of the D3-brane is that it does not couple to the dilaton and 
therefore the equations of motions for the relevant field in the supergravity effective theory 
are simply:
\begin{eqnarray}
&& R_{MN} \;\;\;=\;\;\; T_{MN} \nonumber \\
&& \nabla _M F_{(5)}^{MABCD}=0 \;\;\;\;\;\;  \left(\leftarrow  \;  F^{(5)}_{G_1 \dots G_5}
= \frac{1}{5!} \, \epsilon_{G_1 \dots G_5 H_1 \dots H_5}\,F_{(5)}^{H_1 \dots H_5}\right) \nonumber
\end{eqnarray}
One can make a block-diagonal spherically simmetric ans\"atz for the metric 
$$ds^2 = 
g_{\mu\nu}(x) dx^\mu dx^\nu + g^{(6)}_{ab}(y) dy^a dy^b\;\;\mu ,\nu =0,...,3 \;\; a,b=4,...9
$$ 
choosing in particular the compact components $g_{ab}$ not to depend on the non-compact 
coordinates $x^{\mu}$. Notice that in general the compact components of the metric should depend 
on non-compact coordinates, becoming some of the scalar fields in 4 dimensions. In our case, 
however, we know that in the effective $d=4\;N=2$ supergravity we don't have any vector multiplet 
scalar and the hypers have consistently been put to 0. 

For the 5-form field strenght the ans\"atz is again quite simple because the only supersymmetric 
3-cycles the 3-brane can be wrapped on are the complete holomorphic or anti-holomorphic ones 
because $h_{(2,1)}=0$: 
$$
F_{(5)}(x,y)= F^0_{(2)}(x) \wedge \left(\Omega^{(3,0)} + \bar \Omega^{(0,3)}\right)
$$
With these ans\"atze the equation of motions are satisfied in 10 dimensions and integrating on 
the 6 compact coordinates the lagrangian one gets in 4 dimensions is of a Maxweel-Einstein type 
and the solution turns out to be an extreme R-N solution. The electric and magnetic charges has the following values:
\begin{eqnarray}
e=\frac 12\, \mu \sqrt{V_{D3}^2/{V_{CY}}}\cos \alpha\;,\;
g=\frac 12\, \mu \sqrt{V_{D3}^2/{V_{CY}}}\sin \alpha \nonumber
\end{eqnarray}
where $V_{D3}$ and $V_{CY}$ are the volumes of the 3-brane and of the full compact space 
respectively. The extremality conditions reads: $M^2=\frac 14(e^2+g^2)$, with $M$ the mass of the 
solution. By now $\alpha$ is just an arbitrary parameter. What I will actually show in the following 
is how $\alpha$ depends explicitely on the D3-brane orientation in the compact space. In particular, 
the compactification procedure must preserve Dirac quantization conditions which, being satisfied 
in 10 dimensions with the minimal values of the charges (this result being valid for any for 
D-brane), must still be valid in 4 dimensions.

Let us now move to the microscopic side of the computation and see what happens to 
the previously computed scattering amplitude (\ref{amp10}) once one compactifies. The first thing 
one has to do is to project the 10 dimensional boundary state $|B \rangle$ on its $Z_3$ invariant 
part in this way:
$$
| B_{inv} \rangle = P |B,\theta^a_1 \rangle = \frac 13 (1+g+g^2) |B,\theta^a_1 \rangle 
$$
$P$ is the projection operator and $g$ is the generator of ($2\pi/3$) rotations in 
each of the three planes $X^a,X^{a+1}$ ($a=4,6,8$). 
The net effect is that the boundary state is similar to the previous one but with a 
sum on $\Delta\theta$ angles whose values depend on $g$ (and hence on the particular structure 
of our manifold). In this way:
$$
| B_{inv} \rangle =\frac13 \sum_{\Delta \theta^a} |B,\theta^a_1 +\Delta \theta^a \rangle 
\;\;\;,\;\; \Delta \theta^a = \left(\matrix {0 & 2\pi \over3 & 4\pi \over 3 \cr}
\right)
$$
The full computation gives back the following result for the partition functions, that is 
similar to (\ref{spin}) but with sums on angles $\Delta\theta$: 
\begin{eqnarray}
&& Z_F^{even} \rightarrow 2 \cosh v \sum_{\{\Delta\theta^a\}} 
\prod_a 2 \cos \left(\theta_a + \Delta\theta_a \right) - 2 (2 \cosh 2v + 
\sum_{\{\Delta\theta^a\}} \sum_a 2\cos 2 \left(\theta_a + \Delta\theta_a \right) ) \nonumber \\
&& Z_F^{odd} \rightarrow 2i \sinh v \prod_a 2 \sum_{\{\Delta\theta^a\}} 
\sum_a 2\sin \left(\theta_a + \Delta\theta_a \right) \nonumber
\end{eqnarray}
Using trivial trigonometrical properties of $\Delta\theta_a$ angles which are peculiar, however, 
just of the orbifold, one obtains a very simple result for $Z^{even}_F$ and the 4 dimensional 
amplitude reads:
\begin{equation} 
{\cal A} \sim ( \frac 34 \,\cosh v \; \cos ( \sum_a \theta_a ) \, -\,  \cosh 2v )\, + 0 \nonumber
\end{equation}
In 4 dimensional field theory language the three terms represent respectively the exchange of 
vector fields, of graviton fields and of scalar fields (to which the wrapped D3-brane therefore 
does not couple): the configuration is then of a R-N type (no scalars excited) as we have already 
seen from supergravity side. It is easy to see that there is no contribution from the twisted 
sector because the wrapped D3-brane has mixed boundary conditions in at least one of the three $T^2$ 
componing $T^6/Z_3$ and this is not consistent with twisting.

Notice that in order to find which field the brane couples to in 4 dimensions it would have not 
been necessary to study any scattering amplitude but rather one point functions of supergravity 
fields on the disk representing the D-brane. This is precisely what has been done in \cite{bfis} 
and the results are the same as here. 

As anticipated, it is interesting to see how the D3-brane orientation in the compact space 
affects the values of the electric and magnetic 4 dimensional charges and how Dirac quantization 
condition changes from 10 to 4 dimensions. In dimensions $d=2(q+1)$ Dirac quantization condition 
(Dqc) is: $\, eg'+(-1)^q e'g = 2\pi n$ hence it has a $+$ sign in 10d and a $-$ sign in 4d. 
As it is weel known, D-branes realize Dqc with the minimal amount of charge and for any couple 
of $p$ and $(6-p)$ branes it is always true that $\mu_p\mu_{6-p}\,+\,\mu_{6-p}\mu_p\,=\,4\pi$. 
Once one compactifies the numerical values of the charges is affected by the compact part of the 
bosonic zero-mode part of the boundary state. Indeed the net effect of the compactification is 
that compact space zero-modes aquire a discrete structure.
Before doing any orbifold projection their whole contribution to the scattering 
amplitude turns out to be the following: 
\begin{equation}
\label{muhat}
<\theta^a_1,\vec Y_1|e^{-lH}|\theta^a_2,\vec Y_2>_B = 
\frac{V(B_1) V(B_2)}{\mbox{Vol}(T_6)} = \prod_a 
\frac{|\bar n^{(1)}_a \bar n^{(2)}_{a+1} - \bar n^{(1)}_{a+1} \bar n^{(2)}_a|}
{\sin |\theta_a|}
\end{equation}
where $V(B_1)$ and $V(B_2)$ are the volumes of the two 3-branes and where various 
$\bar n^i_a$ are integers depending explicetly to the geometric configuration of the branes 
in the compact space. The 4 dimensional amplitude therefore becomes the following one:
$${\cal A} = \frac {\hat \mu^2}{\sinh |v|} \int_{0}^{\infty}
\frac {dl}{4\pi l} e^{-\frac {b^2}{4 l}} \frac 1{16} \sum_s Z_B Z^s_F \qquad \mbox{with}
\qquad \hat \mu^2 \equiv \mu^2 \frac{V(B_1) V(B_2)}{\mbox{Vol}(T_6)}
$$
One can now see that Dqc is still valid in 4 dimensions with an integer that depends on 
branes' orientation in the compact space. Moreover, using again trivial trigonometrical 
properties, one can rewrite it as a combinations of 8 different charges (4 electric and 4 
magnetic) whose exact values depend on suitable angles combinations. Using (\ref{odd}) 
and (\ref{muhat}):
\begin{eqnarray}
\label{dirac}
&&\hat \mu^2 \prod_a \sin \theta_a\;=\, 
\sum_{i=1}^4 \left(e^{(1)}_i g^{(2)}_i - g^{(1)}_i e^{(2)}_i 
\right)\;=\; 2 \pi \prod_a |\bar n^{(1)}_a \bar n^{(2)}_{a+1} - 
\bar n^{(1)}_{a+1} \bar n^{(2)}_a|  \nonumber \\
&& e^{(1,2)}_i=\frac{\hat \mu}2 \cos \Phi^{(1,2)}_i \;,\;
g^{(1,2)}_i=\frac{\hat \mu}2 \sin \Phi^{(1,2)}_i \;\;;\;
\Phi^{(1,2)}_i=\theta^4_{(1,2)}\pm\theta^6_{(1,2)}\pm\theta^8_{(1,2)}
\end{eqnarray}
One can see that  there are always at least 4 charges with non vanishing values, independently 
from different values one can choose for the various angles.

By the orbifold projection only one pair of electro-magnetic charges survives, because only one 
angle combination ($\Phi_1=\theta_{(1,2)}^4+\theta_{(1,2)}^6+\theta_{(1,2)}^8$) is $Z_3$ 
invariant and the final result is the following: 
$$
e^{(1,2)}=\frac {\hat \mu}2 \cos \sum_a \theta^a_{1,2} \;\;,\;\;
g^{(1,2)}=\frac {\hat \mu}2 \sin \sum_a \theta^a_{1,2}
$$
still satisfying Dqc. As expected, each 3-brane is a point-like 4 dimensional dyon and couples to 
the unique gauge field present, the graviphoton. Moreover, the previously introduced 
parameter $\alpha$ has now a precise value, being the sum of the 3 angles each brane is tilted 
with respect to the compact space referring directions. For parallel branes one sees 
that $\;e^{(1)}=e^{(2)}\;,\; g^{(1)}=g^{(2)}\;$ and the Lorentz contribution cancels 
(see (\ref{dirac})) as expected for identical dyons in 4 dimensions.

It would be interesting to carry on the orbifold compactification on the type IIA side 
where the number of vector multiplets would be 9 rather than 0 and so there would be non-trivial 
vector multiplet scalars. A very interesting fact is that in this case the $N=2$ supergravity 
effective theory turns out to be a consistent truncation of $N=8$ supergravity and in particular 
the solution one obtains is the most general $1/8$ supersymmetry preserving $N=8$ black hole 
modulo U-duality transformations. The construction of this solution (metric, scalar and vector 
fields) is the main result of a forthcoming paper I'm doing in collaboration with M. Trigiante 
and P. Fr\`e.

Of course, it would also be interesting to consider compactifications on more complicated 
Calabi-Yau spaces on which however a much more involved technique is needed in order to deal 
with D-branes and boundary states in an efficient way. 

\vskip 10pt
\noindent
{\large \bf Acknowledgments}
\vskip 10pt

Work partially supported by EEC contract ERBFMRX-CT96-0045.

\end{document}